\documentclass[trackchanges]{aastex7}
\begin{document}

\title{Stellar Parameters and Orbital Period Estimates for Composite-Spectrum sdB+MS Binaries from LAMOST}

\author[orcid=0000-0003-3832-8864,sname='Li']{Jiangdan Li}
\affiliation{International Centre of Supernovae (ICESUN), Yunnan  Key Laboratory of Supernova Research, Yunnan Observatories, Chinese Academy of Sciences (CAS), Kunming 650216, People's Republic of China}
\email[show]{lijiangdan@ynao.ac.cn}

\author[sname='Xiong']{Jianping Xiong}
\affiliation{International Centre of Supernovae (ICESUN), Yunnan  Key Laboratory of Supernova Research, Yunnan Observatories, Chinese Academy of Sciences (CAS), Kunming 650216, People's Republic of China}
\email{xiongjianping@ynao.ac.cn}

\author[orcid=0000-0002-2577-1990,sname='Li']{Jiao Li}
\affiliation{International Centre of Supernovae (ICESUN), Yunnan  Key Laboratory of Supernova Research, Yunnan Observatories, Chinese Academy of Sciences (CAS), Kunming 650216, People's Republic of China}
\email{lijiao@ynao.ac.cn}

\author{Hai-Liang Chen}
\affiliation{International Centre of Supernovae (ICESUN), Yunnan  Key Laboratory of Supernova Research, Yunnan Observatories, Chinese Academy of Sciences (CAS), Kunming 650216, People's Republic of China}
\affiliation{University of Chinese Academy of Sciences, Beijing 100049, China}
\email{chenhl@ynao.ac.cn}

\author{Hongwei Ge}
\affiliation{International Centre of Supernovae (ICESUN), Yunnan  Key Laboratory of Supernova Research, Yunnan Observatories, Chinese Academy of Sciences (CAS), Kunming 650216, People's Republic of China}
\affiliation{University of Chinese Academy of Sciences, Beijing 100049, China}
\email{gehw@ynao.ac.cn}

\author[sname='Yang']{Mingkuan Yang}
\affiliation{University of Chinese Academy of Sciences, Beijing 100049, China}
\email{chenhl@ynao.ac.cn}
\affiliation{Key Laboratory of Optical Astronomy, National Astronomical Observatories, Chinese Academy of Sciences, Beijing 100101, P.R. China}
\email{yangmk@bao.ac.cn}

\author{Xuefei Chen}
\affiliation{International Centre of Supernovae (ICESUN), Yunnan  Key Laboratory of Supernova Research, Yunnan Observatories, Chinese Academy of Sciences (CAS), Kunming 650216, People's Republic of China}
\affiliation{University of Chinese Academy of Sciences, Beijing 100049, China}
\email{cxf@ynao.ac.cn}

\author{Zhanwen Han}
\affiliation{International Centre of Supernovae (ICESUN), Yunnan  Key Laboratory of Supernova Research, Yunnan Observatories, Chinese Academy of Sciences (CAS), Kunming 650216, People's Republic of China}
\affiliation{University of Chinese Academy of Sciences, Beijing 100049, China}
\email{zhanwenhan@ynao.ac.cn}

\begin{abstract}

Hot subdwarf (sdB) stars in binary systems with main-sequence (MS) companions provide valuable insights into mass transfer and envelope ejection processes in binary evolution. Their mass ratios, orbital periods, and stellar properties encode key information about their evolutionary histories.
In this work, we analyze a sample of 123 composite-spectrum sdB+MS binaries identified from the Large Sky Area Multi-Object Fiber Spectroscopic Telescope Low-Resolution Survey (LAMOST-LRS) Data Release (DR) 8. We adopt atmospheric parameters from spectral decomposition and estimate stellar masses and radii using theoretical evolutionary tracks. Radial velocities for both the hot subdwarfs and cool companions are measured independently through cross-correlation with synthetic templates. Orbital periods are statistically estimated using single-epoch RV separations and a Monte Carlo method that accounts for random inclination and orbital phase.
We find that sdB masses are narrowly distributed around $0.5\,{\rm M}_\odot$, consistent with expectations for core helium-burning stars, while MS companion masses span $0.6$-$1.9\,{\rm M}_\odot$, with most falling between $1.0$ and $1.4\,{\rm M}_\odot$. The inferred orbital-period distribution shows a clear concentration toward long periods, broadly consistent with expectations for binaries formed through stable Roche-lobe overflow. Given that our sample consists of composite-spectrum sdB binaries, mainly sdB+FGK systems, the prevalence of long periods is largely driven by observational selection effects rather than the intrinsic period distribution of the sdB binary population.
This study provides one of the largest uniform catalogs of composite spectrum sdB binaries to date, offering new observational constraints on their physical properties and formation channels.

\end{abstract}



\section{Introduction}\label{sec:intro}

Hot subdwarf (sdB) stars are evolved, compact stars undergoing core helium burning with only very thin hydrogen envelopes \citep{Heber1986, Heber2009, Heber2016}. Located at the blue end of the horizontal branch in the Hertzsprung–Russell diagram, these stars are characterized by high effective temperatures ($T_{\mathrm{eff}} \sim 20{\,}000$--$40{\,}000$\,K)  and surface gravities ($\log g \sim 5.0$--$6.5$) \citep{Heber1986, Heber2009, Heber2016}. Due to their extreme temperatures and compact structure, sdBs contribute significantly to the far-ultraviolet (FUV) output of old stellar populations, and are crucial in explaining the UV upturn observed in elliptical galaxies and bulges of spirals \citep{Greggio1990ApJ, Han2007MNRAS}. Moreover, their relatively simple internal structure makes them excellent targets for asteroseismology \citep[e.g.,][]{Charpinet1996, Fontaine2003} and population studies of Galactic structure and kinematics \citep{Green1986}.

The formation of sdB stars is now widely understood to result from binary evolution \citep{Han2002, Han2003, Han2020}. Binary population synthesis models have identified three principal formation channels: (1) stable Roche-lobe overflow (RLOF), in which the progenitor star loses its hydrogen envelope via stable mass transfer to a companion; (2) common-envelope (CE) ejection, where dynamically unstable mass transfer leads to the formation of a CE and the CE ejection results in the formation of a close binary; and (3) the merger of two helium white dwarfs (He WDs), producing a single sdB star. 
In particular, sdB stars with main-sequence (MS) companions can form through either the stable RLOF or CE channel. Binaries formed through RLOF are expected to have wide orbits, with periods ranging from several hundred to thousands of days, while CE evolution leads to short-period systems, often with orbital periods of less than 10 days \citep{Green2001_sdB, Han2002, Chenxuefei2013MNRAS}. Detailed binary evolution simulations show that the final orbital period and mass ratio of an sdB+MS system are sensitive to initial parameters such as the mass ratio at onset of interaction, orbital separation, and the evolutionary phase of the sdB progenitor. Therefore, observational constraints from well-characterized sdB+MS binaries provide critical tests of binary interaction physics, including CE ejection efficiency, and RLOF stability criteria.

Although the compact sdB binaries from CE evolution have been extensively studied due to their short periods and radial velocity variations, longer-period sdB+MS systems remain less well characterized. These wide systems, often presenting as composite-spectrum binaries, are key to understanding the stable RLOF channel, but they pose significant observational challenges. Since both stellar components contribute appreciably to the observed optical flux, disentangling their properties requires careful spectral decomposition and multi-wavelength modeling.
The first orbits of such long-period systems were published in 2012 \citep{Deca2012MNRAS, Ostensen_Winckel2012_sdB+MS, Vos2012}, marking the beginning of systematic studies of the composite sdB+MS binaries. A major milestone was achieved by \citet{Vos2019MNRAS}, who presented a sample of 23 composite-spectrum sdB+MS binaries with fully determined orbital solutions. These systems, with orbital periods ranging from 1 to 4 years, revealed a striking empirical correlation between orbital period and mass ratio. This trend has significant implications for the long-standing debate on RLOF stability and mass transfer physics. However, the sample also exhibited unexpected features—most notably, a prevalence of eccentric orbits in systems that should have undergone tidal circularization prior to mass transfer. Various mechanisms have been proposed to explain the observed orbital properties, including interactions with circumbinary disks \citep{Vos2015AA}. Nevertheless, the origin of the eccentricities observed in long-period sdB systems remains unclear. Furthermore, a possible bimodality in the period–mass ratio and period–eccentricity relations suggests a complex and potentially diverse set of formation histories.

Despite the progress made, wide sdB+MS binaries are still underrepresented in large-scale surveys. This is due in part to the difficulty of identifying and analyzing composite-spectrum systems, which require careful treatment of both hot and cool components. High-resolution spectroscopic studies have been limited to small samples, with 24 systems decomposed from the SPY survey \citep{Lisker2005A&A}, 29 from GALEX \citep{Nemeth2012MNRAS_sdB}, and only 9 high-quality decompositions from VLT/UVES spectra reported by \citet{Vos2018MNRAS}. More recently, \citet{Nemeth2021_SB744} performed a detailed analysis of SB 744, a long-period sdOB+G1V binary, revealing extreme abundance patterns that indicate a Population II origin. These findings underscore the scientific value of such systems, and highlight the need for larger, homogeneous samples.

To address the need for larger and more homogeneous samples of hot subdwarf binaries, we analyze a subset of composite-spectrum sdB+MS systems identified by \citet{LeiZhenxin2023ApJ} from the the Large Sky Area Multi-Object Fiber Spectroscopic Telescope (LAMOST) low-resolution survey (LRS) data release (DR) 8. This catalog includes 131 binaries with atmospheric parameters for both components derived through spectral decomposition. The wide sky coverage, moderate spectral resolution ($R \sim 1{\,}800$), and high multiplexing capability of LAMOST make it well suited for constructing statistically meaningful samples of rare binary populations. 
In this work, we select a refined subset of high-quality systems with reliable Gaia parallaxes. Our primary goal is to determine the fundamental parameters of both components, including effective temperature, surface gravity, mass, and radius, using atmospheric measurements in combination with theoretical evolutionary tracks.
This paper is structured as follows. Section~\ref{sec:sample} describes the sample selection from the LAMOST survey and outlines the criteria for identifying composite-spectrum binaries. Section~\ref{sec:parameters} details the determination of atmospheric parameters and the use of theoretical evolutionary tracks to estimate stellar masses and radii. Section~\ref{sec:orbit} outlines the statistical approach for estimating orbital periods, while Section~\ref{sec:results} presents the resulting parameter distributions and empirical correlations. In Section~\ref{sec:discussion}, we discuss the implications of these findings for binary formation and evolution, and compare our results with theoretical population synthesis predictions. A summary of our main conclusions is given in Section~\ref{sec:summary}.

\section{Sample Selection} \label{sec:sample}

The sample analyzed in this study is based on the catalog of composite-spectrum sdB binaries presented by \citet{LeiZhenxin2023ApJ}, which was constructed using spectroscopic data from LAMOST-LRS DR 8. This catalog contains 131 systems in which both a hot subdwarf and a cool main-sequence companion are spectroscopically identified via composite features in low-resolution spectra.

LAMOST is a 4-meter quasi-meridian reflecting Schmidt telescope optimized for wide-field, high-throughput spectroscopic surveys \citep{Cuixiangqun2012RAA, Zhaogang2012RAA}. Its low-resolution mode provides spectral coverage across the 3\,700–9\,000\,\AA\ range at a spectral resolution of $R \sim 1\,800$. With 4\,000 optical fibers distributed across a 5-degree field of view, LAMOST can observe thousands of stars simultaneously, making it particularly effective for identifying composite hot subdwarf binaries across the sky.

To construct a reliable sample of sdB+MS binaries, \citet{LeiZhenxin2023ApJ} applied a series of stringent selection criteria. Only spectra with a signal-to-noise ratio (SNR) greater than 10 in the $u$-band were considered to ensure sufficient quality for detecting features from both stellar components. Each selected spectrum was required to show evidence of a hot subdwarf through broad hydrogen Balmer and He I absorption lines, and a cool companion via metal absorption lines such as the Mg I triplet ($\lambda\lambda$5\,167, 5\,173, 5\,184) and the Ca II infrared triplet ($\lambda\lambda$8\,500, 8\,544, and 8\,664). Systems with ambiguous or marginal composite features were excluded, ensuring that both stellar components were robustly detected and spectroscopically separable. 

To ensure the reliability of the derived stellar properties, we applied additional quality criteria to refine the initial DR8-based sample. All systems were cross-matched with Gaia DR3 \citep{GaiaDR3_2023A&A}, and distances were adopted from \citet{GaiaEdr3_distance2021AJ}. More than 94\% of the sample has distance uncertainties below 20\%, and over 98\% lie within 5\,000 pc. Accurate distance measurements are essential for population studies and for deriving physical parameters such as stellar luminosity and radius.
Given that the composite spectra include contributions from both components, different spectral regions are dominated by different stars. In the blue (wavelengths shorter than $\sim$5\,000\,\AA), the flux is dominated by the hot subdwarf, allowing robust determination of its atmospheric parameters and radial velocity. In contrast, the red part of the spectrum (wavelengths longer than $\sim$8\,000\,\AA) is dominated by the main-sequence companion, particularly through the Ca~II infrared triplet. To ensure reliable parameter measurements for both components, we required each spectrum to have a SNR greater than 10 in both the $g$-band (for H and He lines in the sdB) and the $i$-band (for the Ca~II features of the MS companion).

After applying these filters, the final sample includes 123 composite-spectrum sdB+MS binary candidates. Each system has high-quality LAMOST spectra, well-constrained atmospheric parameters for the sdB component, and reliable Gaia astrometry. For a subset of systems, multi-epoch spectra from LAMOST DR 12 are available, enabling further analysis of radial velocities. This sample represents one of the largest homogeneous collections of composite hot subdwarfs identified via spectroscopic decomposition in a wide-field survey.

Although the composite-spectrum sdBs identified by \citet{LeiZhenxin2023ApJ} were classified as binary candidates, we have quantitatively assessed the probability of chance alignments using Gaia~DR3 data. To appear as a composite spectrum in the optical, the flux contributions from the hot subdwarf and the cool companion must be comparable within the LAMOST wavelength range. In practice, this requires that the Gaia $G_{\mathrm{RP}}$ magnitude, which includes light from both components, does not differ significantly from the $G_{\mathrm{BP}}$ magnitude. In our sample, the observed difference between $G_{\mathrm{RP}}$ and $G_{\mathrm{BP}}$ ranges from $-0.79$ to $+0.42$\,mag, confirming that the flux contributions of the two components are indeed comparable. Therefore, when estimating the probability of chance alignments, we adopted a conservative brightness range of $\pm1$\,mag around the expected companion magnitude, which comfortably encompasses all observed cases.
Based on this constraint, we estimated the probability of chance alignments by counting nearby sources within $0.25^\circ$ of each sdB and within $\pm1$\,mag of the expected companion brightness in Gaia~DR3 \citep{GaiaDR3_2023A&A}. Assuming a uniform distribution of field stars in solid angle, we converted these counts to local stellar surface densities and applied Poisson statistics for a 1.65$^{\prime\prime}$-radius LAMOST fiber \citep{Cuixiangqun2012RAA, Zhaogang2012RAA}. The summed contamination probability across all 123 systems is $N_{\mathrm{false}} = 0.1$, corresponding to a mean probability of an optical double of $P \sim \rho \pi R_{\rm ang}^2 = 8.7\times10^{-4}$ per target. Therefore, the likelihood that our sample is significantly contaminated by chance projections is negligible, and the vast majority of systems are highly probable to be genuine binaries. Future time-series spectroscopy or multi-epoch Gaia astrometry will be valuable for confirming binarity in the remaining candidates.

\section{Atmospheric and Orbital Parameters} \label{sec:parameters}
\subsection{Atmospheric Parameters from Spectral Decomposition}

Atmospheric parameters for both stellar components were adopted from the spectral decomposition analysis of \citet{LeiZhenxin2023ApJ}, who modeled the LAMOST DR8 composite spectra using the XTGRID fitting code \citep{Nemeth2012MNRAS_sdB}. The sdB parameters (\( T_{\rm eff, sdB} \), \( \log g_{\rm sdB} \), and \( \log y \)) were primarily constrained by hydrogen and helium line profiles, with typical uncertainties of about 515\,K in \( T_{\rm eff, sdB} \) and 0.08\,dex in \( \log g_{\rm sdB} \). The atmospheric parameters of the MS companions are less precisely constrained because their flux contribution is relatively weak. \citet{LeiZhenxin2023ApJ} reported the best-fit values of \( T_{\rm eff, MS} \) and \( \log g_{\rm MS} \) without formal uncertainties, as the low-resolution LAMOST spectra are not highly sensitive to gravity-dependent line shapes. In this study, we use the sdB parameters and \( T_{\rm eff, MS} \) values from their analysis, while refining the surface gravities and other evolutionary parameters of the MS components independently using stellar evolution models, as described in the following section.

\subsection{sdB Masses and Radii from Evolutionary Tracks}

To estimate the fundamental parameters of the sdB components in our sample, we adopt the method of \citet{zhangxianfei2009}, which relates the most probable mass of a core helium-burning sdB star to the observable quantity \(\log(T_{\rm eff}^4 / g)\). We note that this most probable mass corresponds to the sdB core, which dominates the total mass due to the extremely thin envelope, typically with masses $M_{\rm env} \lesssim 0.01$\,M$_\odot$ \citep{Heber2009,Heber2016}. For practical purposes, the core mass is assumed to represent the total stellar mass, providing a lower limit to both the mass and the corresponding radius.
This approach relies on the physical connection between surface gravity \( g \), stellar radius \( R \), and effective temperature \( T_{\rm eff} \). Since luminosity scales as \( L \propto T_{\rm eff}^4 R^2 \) and surface gravity as \( g \propto M / R^2 \), the ratio \( T_{\rm eff}^4 / g \) effectively traces the luminosity-to-mass ratio of the star. \citet{zhangxianfei2009} constructed a dense grid of evolutionary models for hot subdwarfs with core masses ranging from 0.33 to 1.4\,\( {\rm M}_\odot \) at solar metallicity (\( Z = 0.02 \)), allowing them to empirically map \( \log(T_{\rm eff}^4 / g) \) to the most probable mass \( M_{\rm sdB} \) during the core helium-burning phase.

For each system in our sample, we compute \( \log(T_{\rm eff}^4 / g) \) using the spectroscopically determined values of \( T_{\rm eff, sdB} \) and \( \log g_{\rm sdB} \) from \citet{LeiZhenxin2023ApJ}, and interpolate the empirical relation from \citet{zhangxianfei2009} to obtain the most probable sdB mass \( M_{\rm sdB} \). The radius of each sdB component is then calculated using the standard relation
\begin{equation}
R_{\rm sdB} = \sqrt{\frac{G M_{\rm sdB}}{g_{\rm sdB}}},
\end{equation}
where \( G \) is the gravitational constant. 
Uncertainties on the sdB mass and radius were estimated via Monte Carlo sampling of the spectroscopic parameters. For each star, we generated $10^4$ realizations of $(T_{\rm eff, sdB},\log g_{\rm sdB})$ assuming Gaussian errors from \citet{LeiZhenxin2023ApJ}, computed the corresponding $M_p$ for each realization, and adopted the median of the resulting distribution as the best estimate. The 16th and 84th percentiles define the $1\sigma$ credible interval, and the mean Monte Carlo-derived mass uncertainty across our sample is 0.06\,M$_\odot$.
This procedure yields physically consistent sdB masses and radii, generally within the canonical range of core helium-burning stars (\(\sim0.4\text{--}0.6\,{\rm M}_\odot\); e.g., \citet{Heber2009, Heber2016}), while properly accounting for observational uncertainties in $T_{\rm eff}$ and $\log g$.

\subsection{Initial Estimates from ZAMS Evolutionary Models}

To estimate the fundamental parameters of the MS companions, we adopt solar-metallicity ($Z = 0.02$) zero-age main-sequence (ZAMS) models from the PARSEC v1.2S evolutionary tracks \citep{Bressan2012MNRAS_PARSEC, Chenyang2014MNRAS_PARSEC}.  
The assumption that the main-sequence companion lies on the ZAMS is well justified in the context of hot subdwarf binaries. The core helium-burning phase of an sdB star lasts approximately $\sim 10^8$\,yr \citep{Han2002, Han2003, Arancibia-Rojas2024MNRAS_sdB_MESA}, while MS stars with masses between 0.8 and 2.5\,${\rm M}_\odot$ have lifetimes exceeding $10^9$\,yr at solar metallicity \citep{Bressan2012MNRAS_PARSEC, Chenyang2014MNRAS_PARSEC}.  
In the CE formation channel, the system initially consists of two main-sequence stars; the more massive star evolves first, fills its Roche lobe, and the subsequent CE ejection leaves behind an sdB star. The less massive companion, still on the main sequence, evolves on a much longer timescale, making the ZAMS approximation valid.  
In the stable RLOF channel, the initially more massive star transfers mass to its companion during the RGB phase, and the accretion rejuvenates the secondary, resetting its evolutionary clock and producing a structure closely resembling a ZAMS star \citep{Han2002, Han2003}.  
As such, the MS companions are expected to remain close to the ZAMS throughout the sdB's lifetime. Furthermore, few spectroscopic signatures of post-main-sequence evolution (e.g., enhanced line broadening, luminosity class indicators) are detected in the sample, supporting the ZAMS assumption \citep{LeiZhenxin2023ApJ}.

We adopt the effective temperatures of the MS components ($T_{\rm eff,MS}$) derived by \citet{LeiZhenxin2023ApJ} through spectral decomposition as the primary input.  
Formal uncertainties for the atmospheric parameters of the MS companions were not provided by \citet{LeiZhenxin2023ApJ}.  
To obtain realistic estimates, we empirically derived uncertainties from the observed dispersion of fitted parameters across our sample.  
The standard deviation of $T_{\rm eff,MS}$ values, $\sigma(T_{\rm eff,MS}) = 779$\,K, was adopted as the representative $1\sigma$ uncertainty.  
This approach better captures the systematic effects inherent to low-resolution composite spectra and provides a more conservative and realistic error estimate.  

For each system, we generate 1\,000 random realizations of $T_{\rm eff,MS}$ drawn from this Gaussian distribution and map each realization to the solar-metallicity ($Z = 0.02$) PARSEC ZAMS models to interpolate the corresponding stellar mass ($M_{\rm MS}$), radius ($R_{\rm MS}$), and surface gravity ($\log g_{\rm MS}$). 
The resulting distributions naturally account for the propagation of the temperature uncertainty through the evolutionary models.  
Small changes in metallicity have negligible influence on these parameters, as the adopted uncertainty in $T_{\rm eff,MS}$ (779\,K) dominates over the variation introduced by plausible metallicity differences \citep{Bressan2012MNRAS_PARSEC, Chenyang2014MNRAS_PARSEC}.  
For each system, we adopt the median of the resulting distributions as the best estimates of $M_{\rm MS}$, $R_{\rm MS}$, and $\log g_{\rm MS}$, and use the standard deviation as the corresponding $1\sigma$ uncertainty. The mean uncertainties are $\sigma(M_{\rm MS}) = 0.23\,\rm M_\odot$ and $\sigma(R_{\rm MS}) = 0.22\,\rm R_\odot$.

These model-derived parameters form the basis for estimating the mass ratios of the systems and for investigating their correlations with orbital periods. By relying on well-tested stellar evolutionary tracks and physically motivated assumptions about the companions' evolutionary state, this method provides consistent and realistic initial conditions for population-level analysis of sdB+MS binaries.

\section{Radial Velocities and Orbital Period Estimation} \label{sec:orbit}

\subsection{Radial Velocity Measurements} \label{sec:rv_measurements}

To measure the radial velocities (RVs) of both components in the composite-spectrum sdB+MS binaries, we employed a cross-correlation technique using synthetic spectral templates representative of each stellar type. Given the limited number of multi-epoch observations for most systems, our primary goal is to characterize the RVs of both the hot subdwarf and the cool main-sequence companion from individual spectra and assess their instantaneous RV separations.

Synthetic template spectra were generated separately for each component. For the sdB stars, we adopted typical atmospheric parameters of \( T_{\rm eff, sdB} = 35{\,}000 \)\,K and \( \log g_{\rm sdB} = 5.0 \)\,dex \citep{Heber2016}, selecting synthetic spectra from the TLUSTY grid \citep{Lanz2003, Hubeny2017_TLUSTY} and convolving them to match the resolution of the LAMOST low-resolution spectra (\( R \sim 1{\,}800 \)). The cross-correlation was performed over the 4\,000–5\,000\,\AA\ range, which contains strong Balmer lines (H$_\delta$, H$_\gamma$, H$_\beta$) and He\,I absorption lines. In this region, the flux is dominated by the sdB component, allowing for reliable RV extraction.
For the MS companions, we adopted a representative template with \( T_{\rm eff, MS} = 6{\,}000 \)\,K and \( \log g_{\rm MS} = 4.5 \)\,dex, typical of mid-F to early-G type stars \citep{Gray2009book}. These spectra were selected from the Kurucz model grid \citep{Kurucz1984, Kurucz1993} and similarly convolved to match LAMOST resolution. The RVs of the companions were measured over the 8\,450–8\,700\,\AA\ range, which includes the Ca\,II infrared triplet (8\,500\,\AA, 8\,544\,\AA, and 8\,664\,\AA; vacuum wavelengths). These features are strong in cool stars and largely uncontaminated by sdB flux, providing robust tracers for the companion's motion.

For each LAMOST spectrum from DR 8 or DR 12 with a signal-to-noise ratio greater than 10, we extracted two wavelength segments corresponding to the sdB and MS template regions. RVs were measured by cross-correlating each segment with the corresponding synthetic template using the cross-correlation function (CCF) module in the {\tt laspec} software package \citep{Zhangbo2020ApJS, Zhangbo2021ApJS}. Each spectrum thus yields two RV measurements: one from the Balmer and He\,I lines of the sdB, and one from the Ca\,II triplet of the MS companion. All spectra were corrected to the barycentric frame prior to RV analysis.

We successfully obtained RVs for both components in all 123 systems. Of these, 57 have only a single LAMOST observation, 37 have two epochs, and 29 have three or more. One system has as many as 14 spectroscopic epochs. However, most multi-epoch observations lack sufficient phase coverage for fitting complete orbital solutions. The cadence is typically sparse and irregular, so we do not attempt to derive full RV curves or orbital periods from these data.

\subsection{Orbital Period Estimation}
\label{sec:orbital_period}

To estimate the orbital periods of the composite-spectrum sdB+MS binaries in our sample, we use single-epoch RV measurements derived from LAMOST-LRS spectra. As described in Section~\ref{sec:rv_measurements}, the RVs of both the sdB and MS components are independently measured via cross-correlation with synthetic templates. The instantaneous RV separation for each system is defined as
\begin{equation}
\Delta {RV} = | {RV}_{\mathrm{sdB}} - {RV}_{\mathrm{MS}} |.
\end{equation}
Due to the limited number of spectroscopic epochs for most systems, full orbital solutions cannot be determined. Instead, we estimate orbital periods using a statistical approach based on simplified binary dynamics and Monte Carlo (MC) simulations.

\subsubsection{Monte Carlo Inference} \label{subsubsec:MC}

Hot subdwarf binaries with MS companions are believed to originate from two primary evolutionary channels. The first is the CE ejection channel, which produces short-period systems with \( P_{\rm orb} \lesssim 10 \) days. The second is the stable RLOF channel, leading to wider binaries with orbital periods typically ranging from several hundred to over 1\,600 days \citep{Han2002, Han2003, Chenxuefei2013MNRAS}. In both scenarios, tidal interactions during the mass-transfer phase are expected to efficiently circularize the orbits \citep{Chenxuefei2013MNRAS}.
However, recent observational studies challenge this assumption. \citet{Vos2019MNRAS} found that many wide sdB+MS systems exhibit significant eccentricities, possibly due to interactions with circumbinary disks \citep{Vos2015AA}.     
Given the lack of multi-epoch RV coverage in our data, orbital eccentricities cannot be directly constrained. We therefore adopt circular orbits as a simplifying assumption for the baseline period estimation, while explicitly assessing the impact of eccentricity through forward-modeling tests
(Section~\ref{sec:period_validation}).

Under the circular orbit approximation, the sum of the RV semi-amplitudes ($K_{\mathrm{sdB}} + K_{\mathrm{MS}}$) can be approximated by the observed RV separation corrected for orbital phase:
\begin{equation}
K_{\mathrm{sdB}} + K_{\mathrm{MS}} = \frac{\Delta {RV}}{\sin \tau},
\label{equa:deltaRV}
\end{equation}
where $\tau$ is the orbital phase at the time of observation. The orbital period is then derived from Kepler's third law:
\begin{equation}
P_{\rm orb} = \frac{2\pi G}{(K_{\mathrm{sdB}} + K_{\mathrm{MS}})^3} \left(M_{\mathrm{sdB}} + M_{\mathrm{MS}}\right) \sin^3 i,
\label{equa:Porb}
\end{equation}
where $M_{\mathrm{sdB}}$ and $M_{\mathrm{MS}}$ are the component masses, and $i$ is the orbital inclination angle.
Since both the orbital inclination (\(i\)) and phase (\(\tau\)) are unknown, we performed a Monte Carlo simulation to explore their plausible distributions.
For each system, we generated 1\,000 realizations by sampling \(i\) from a distribution proportional to \(\sin i\), which represents the expected orientation distribution for randomly aligned binary systems. The orbital phase \(\tau\) was drawn uniformly from the range \([0, \pi]\) radians. For each realization, we computed the radial velocity amplitudes and the corresponding orbital period using Equations~(\ref{equa:deltaRV}) and~(\ref{equa:Porb}). Measurement uncertainties in the radial velocities were incorporated through Gaussian perturbations. The final orbital period for each system was taken as the median value from the simulated period distribution, with the 16th and 84th percentiles representing the uncertainty range.
This procedure yields a broad probability distribution of possible orbital periods for each system rather than a uniquely constrained value. Accordingly, the median period of an individual system should be regarded only as a descriptive statistic of a highly degenerate distribution, not
as a physically well-determined orbital solution. Although individual orbital periods are poorly constrained, the ensemble of MC realizations across the full sample provides statistical information on the population-level orbital period distribution. All population-level results presented below are therefore based on the combined MC samples of all systems, rather than on single representative values per binary.

\begin{figure}[ht!]
\centering
\includegraphics[width=0.9\textwidth]{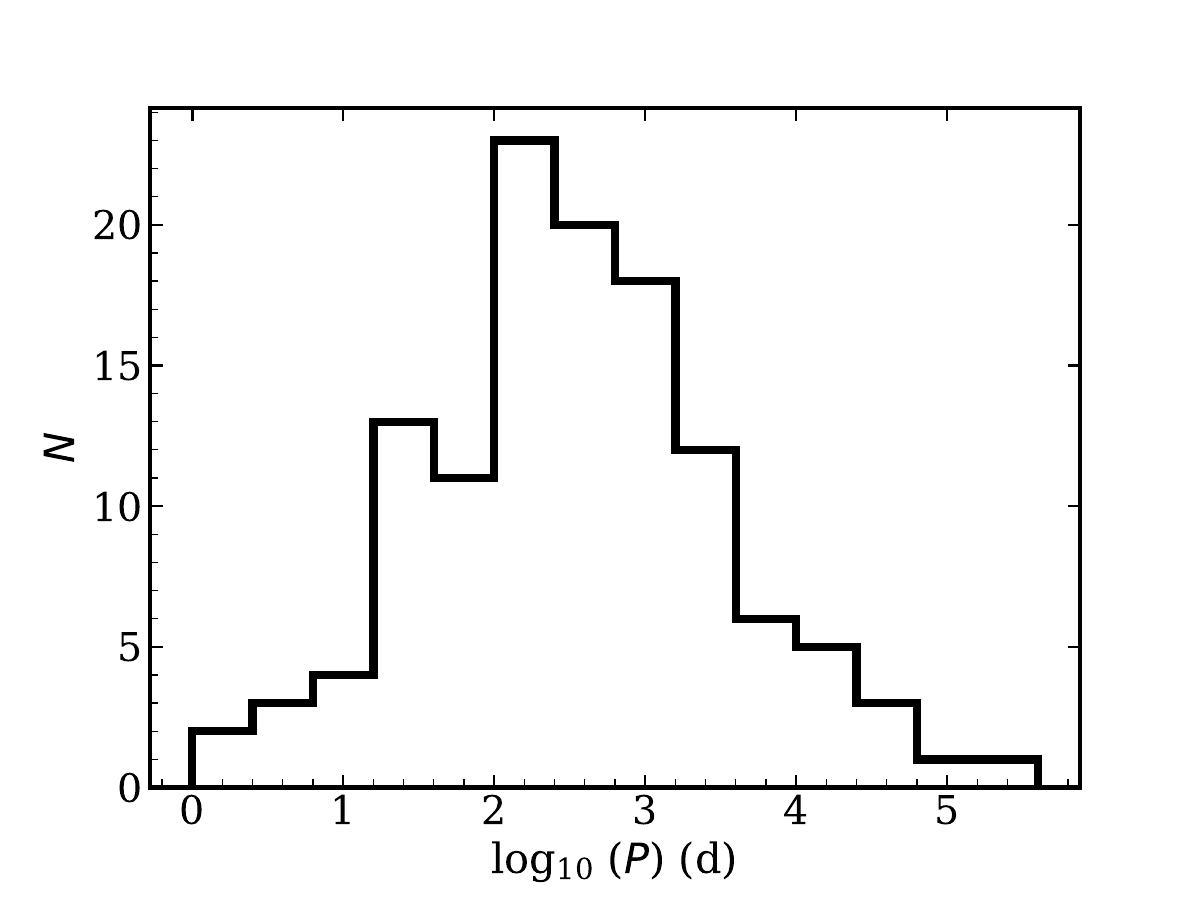}
\caption{
Distribution of inferred orbital periods for the 123 composite-spectrum sdB+MS systems.
Individual orbital periods are not constrained due to strong degeneracies inherent to single-epoch RV measurements. The distribution indicates that most realizations occupy the long-period regime, extending roughly from a few tens to several thousand days, broadly consistent with expectations for stable Roche-lobe overflow.
}
\label{fig:orbital_period_distribution}
\end{figure}

Among the 123 systems with measured radial velocity separations and estimated component masses, the resulting orbital period distribution covers a wide range, from a few days to more than $10^4$ days (Figure~\ref{fig:orbital_period_distribution}). 
Extremely long inferred periods arise naturally from the strong dependence of the inferred period on orbital phase and inclination, particularly for observations obtained near conjunction. Such values should therefore not be interpreted as evidence for intrinsically ultra-wide binaries.
Despite these limitations, the inferred distribution shows a broad concentration toward long orbital periods, extending from a few tens to several thousand days, consistent with expectations for binaries formed through stable RLOF.
These results indicate the statistical presence of the formation channels in the sample, while precluding evolutionary classification of individual systems.

\subsubsection{Validation via Forward Modeling}
\label{sec:period_validation}

To assess the robustness of the inferred orbital-period distributions and to quantify potential biases introduced by unknown orbital parameters, we performed forward-modeling tests using synthetic sdB+MS binary populations. These synthetic systems were subjected to the same single-epoch RV sampling and Monte Carlo inversion procedure as applied to the observed sample.

To ensure full methodological consistency, the synthetic populations were constructed to closely mirror the assumptions adopted in the observational analysis. Each simulation consists of 10\,000 synthetic sdB+MS binaries, substantially exceeding the observed sample size (123 systems) in order to reduce Monte Carlo noise and stabilize the inferred population-level distributions. Intrinsic orbital periods were drawn either from a log-uniform distribution between 100 and 10\,000~days or from population-synthesis period distributions taken from \citet{Chenxuefei2013MNRAS} (their Figure~2) and \citet{Rodriguez-Segovia2025MNRAS_sdBpopulation} (their Figure~8). The latter corresponds to sdB binaries with early-type main-sequence companions ($M \gtrsim 1\,{\rm M}_\odot$) and spans the dominant mass range of our sample. Component masses were sampled independently, with sdB masses uniformly distributed between 0.3 and 0.9~${\rm M}_\odot$ and main-sequence companion masses between 0.65 and 1.7~${\rm M}_\odot$, consistent with the ranges inferred from spectral decomposition. Orbital inclinations were drawn assuming isotropic orientations, and orbital phases were sampled uniformly in $[0,\pi]$. For simulations including eccentric orbits, eccentricities were drawn uniformly between $e=0$ and $e=1$, with randomized arguments of periastron.

\begin{figure}
\centering
\includegraphics[width=0.65\columnwidth]{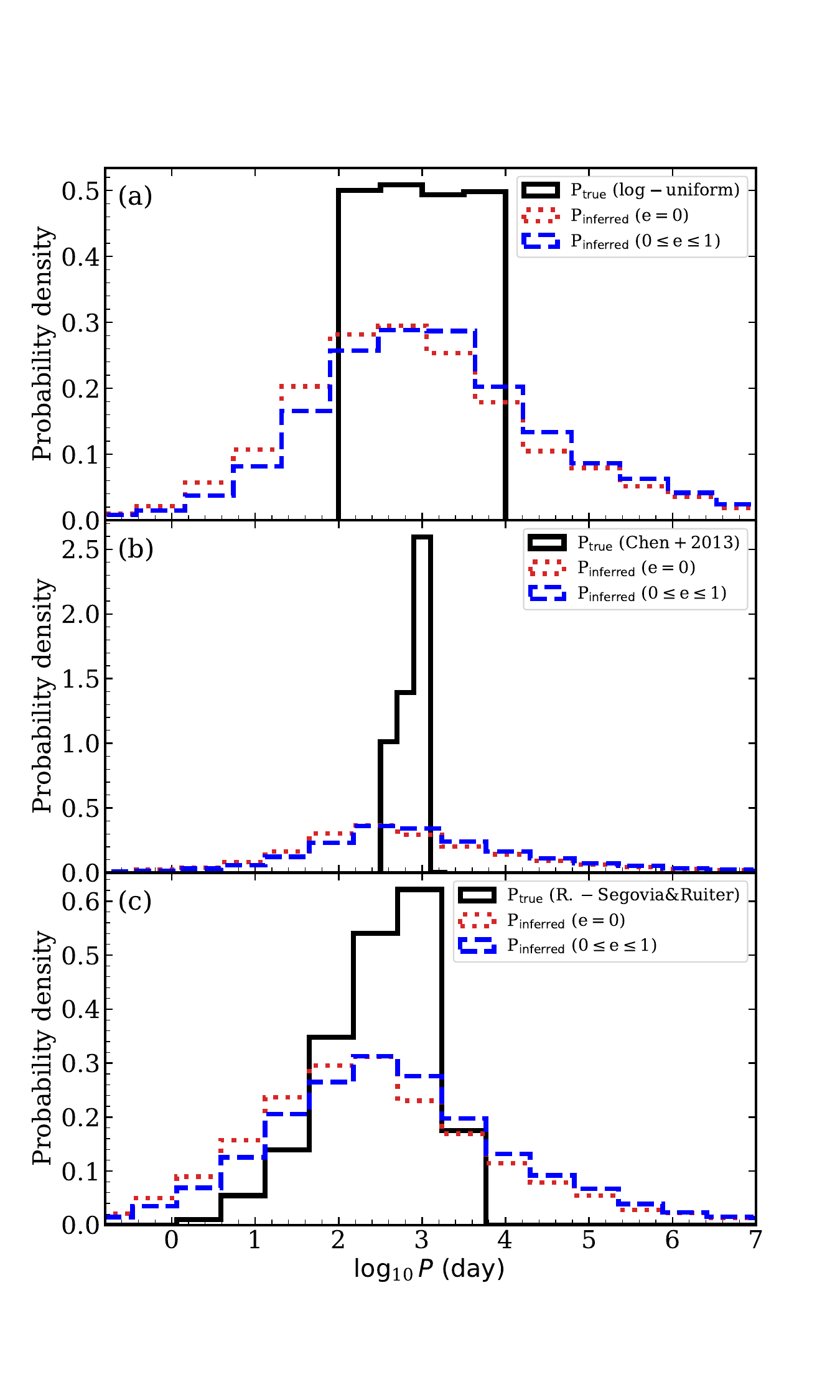}
\caption{
Comparison between intrinsic orbital-period distributions and the corresponding inferred period distributions obtained from the single-epoch Monte Carlo inversion. The black histogram shows the true input period distribution of the synthetic population. The colored histograms show the inferred period distributions assuming circular orbits ($e=0$; red dotted lines) and eccentric orbits with eccentricities drawn uniformly between $e=0$ and $e=1$ (blue dashed lines). Panel (a) uses a log-uniform intrinsic period distribution between 100 and 10\,000~days. Panels (b) and (c) adopt intrinsic period distributions from binary population-synthesis models of \citet{Chenxuefei2013MNRAS} and \citet{Rodriguez-Segovia2025MNRAS_sdBpopulation}, respectively.
}
\label{fig:period_validation}
\end{figure}

For each synthetic system, a single-epoch RV difference was computed and the orbital period was inferred using the same Monte Carlo inversion procedure applied to the observed sample (Section~\ref{subsubsec:MC}), with 1\,000 realizations per system. This guarantees a direct one-to-one correspondence between the forward-modeling and inverse-inference steps. The resulting comparisons between intrinsic and inferred period distributions are shown in Figure~\ref{fig:period_validation} for the log-uniform case, the \citet{Chenxuefei2013MNRAS} distribution, and the \citet{Rodriguez-Segovia2025MNRAS_sdBpopulation} distribution, respectively, under both circular and eccentric-orbit assumptions.
In both panels, the inferred period distributions exhibit extended short- and long-period tails, reflecting the intrinsic degeneracy of single-epoch RV measurements with respect to orbital inclination, phase, and eccentricity. Nevertheless, the dominant period regime at $P \sim 10^{2}$–$10^{3}$~days is robustly recovered and remains insensitive to the assumed intrinsic period prior or eccentricity distribution. 
Adopting circular orbits or allowing for orbital eccentricity does not significantly shift the location of the main peak nor alter the dominance of the long-period population. Because the eccentricity distribution of the observed systems is not constrained by the available data, we adopt the circular-orbit assumption when inferring the orbital-period distribution for the LAMOST sample, treating it as a simplifying approximation rather than a physical description of individual systems.

The mapping between intrinsic orbital periods and single-epoch inferred periods is highly degenerate, rendering the inverse problem fundamentally ill-posed. We therefore do not attempt to deconvolve the observed period distribution to recover the intrinsic one. Instead, we rely on forward modeling to quantify statistical biases and to assess the robustness of population-level conclusions. Although the inferred orbital periods span a wide range (Figure~\ref{fig:orbital_period_distribution}), the population-level distribution shows a clear concentration toward long-period systems extending from a few tens to several thousand days. The forward-modeling tests demonstrate that this broad long-period regime is robust against the uncertainties inherent in single-epoch radial velocity measurements. This dominance arises naturally from the selection bias inherent to composite-spectrum sdB binaries, primarily sdB+FGK systems, which are detectable only when both components contribute appreciably to the optical spectrum. Such systems preferentially trace long-period configurations produced through stable RLOF, whereas short-period post-CE binaries often host low-luminosity companions (e.g., white dwarfs or M dwarfs) that are not detectable in combined spectra (see Figures~13-16 in \citealt{Han2003}). Nevertheless, at the population level the inferred distribution is dominated by systems consistent with post-RLOF evolution, providing statistical constraints on the evolutionary pathways of composite-spectrum sdB+MS binaries.

\subsection{Summary of Stellar and Orbital Parameters}
The refined sample of 123 composite-spectrum sdB+MS binaries includes reliable spectroscopic parameters, estimated component masses and radii, single-epoch radial velocities, and statistically inferred orbital periods. This catalog, summarized in Table~\ref{tab:parameter_summary}, provides a homogeneous dataset for investigating the physical properties and binary evolution pathways of hot subdwarf systems.

\begin{table}[htbp]
    \centering
    \caption{Stellar and orbital parameters of the composite sdB+MS binary sample.}
    \label{tab:parameter_summary}
    \begin{tabular}{l c c c c c}
    \hline
Num  &  Label                  & Units       &  Definitions \\
\hline
1    &  RA                     & deg         &  Right ascension at epoch 2000.0 (ICRS) \\
2    &  Dec                    & deg         &  Declination at epoch 2000.0 (ICRS)\\
3    & $T_{\rm eff,sdB}$       & K           & Effective temperature of sdB  \\
4    & e\_$T_{\rm eff, sdB}$   & K           & Standard error in $T_{\rm eff,sdB}$ \\
5    & $\log{g_{\rm sdB}}$     & dex         & Surface gravity of sdB\\
6    & e\_$\log{g_{\rm sdB}}$  & dex         & Standard error in $\log{g_{\rm sdB}}$ \\
7    & $T_{\rm eff,MS} $       & K           & Effective temperature of MS  \\
8    & e\_$T_{\rm eff,MS}$     & K           & Standard error in $T_{\rm eff,MS}$ \\
9   & $\log{g_{\rm MS}}$      & dex         & Surface gravity of MS\\
10   & e\_$\log{g_{\rm MS}}$   & dex         & Standard error in $\log{g_{\rm MS}}$ \\
11   & $M_{\rm sdB}$           & M$_\odot$   & Mass of sdB  \\
12   & e\_$M_{\rm sdB}$        & M$_\odot$   & Standard error in $M_{\rm sdB}$ \\
13   & $R_{\rm sdB}$           & R$_\odot$   & Radius of sdB  \\
14   & e\_$R_{\rm sdB}$        & R$_\odot$   & Standard error in $R_{\rm sdB}$ \\
15   & $M_{\rm MS}$            & M$_\odot$   & Mass of MS  \\
16   & e\_$M_{\rm MS}$         & M$_\odot$   & Standard error in $M_{\rm MS}$ \\
17   & $R_{\rm MS}$            & R$_\odot$   & Radius of MS  \\
18   & e\_$R_{\rm MS}$         & R$_\odot$   & Standard error in $R_{\rm MS}$ \\
19   & $RV_{\rm sdB}$          & km/s        & Radial velocity of sdB \\
20   & e\_$RV_{\rm sdB}$       & km/s        & Standard error in $RV1$ \\
21   & $RV_{\rm MS}$           & km/s        & Radial velocity of MS \\
22   & e\_$RV_{\rm MS}$        & km/s        & Standard error in $RV2$ \\
23   & $P$                     & day         & Calculated period \\
24   & e\_$P$                  & day         & Standard error in $P$ \\
25   & Dist                    & pc          & Distance \\
26   & e\_Dist                 & pc          & Standard error in distance \\
27   & $E(B-V)_{\rm 3D}$       & mag         & Reddening from 3D dustmap\\

%30  & CE\_candidates          &             & Y = Yes; N = No \\
%\enddata
\hline

\end{tabular}
\vspace{0.5em}
    \begin{minipage}{0.9\textwidth}
    \footnotesize
    \textbf{Note.} $T_{\rm eff,sdB}$, $\log{g_{\rm sdB}}$, $T_{\rm eff,MS} $, $\log{g_{\rm MS}}$ and their errors are taken from \citet{LeiZhenxin2023ApJ}. 
    Distance is adopted from Gaia eDR3 distances \citep{GaiaEdr3_distance2021AJ}.
    \end{minipage}
\end{table}

\section{Results} \label{sec:results}

\subsection{Mass Distributions of sdB and MS Components}
\begin{figure}
\centering
\includegraphics[width=0.9\textwidth]{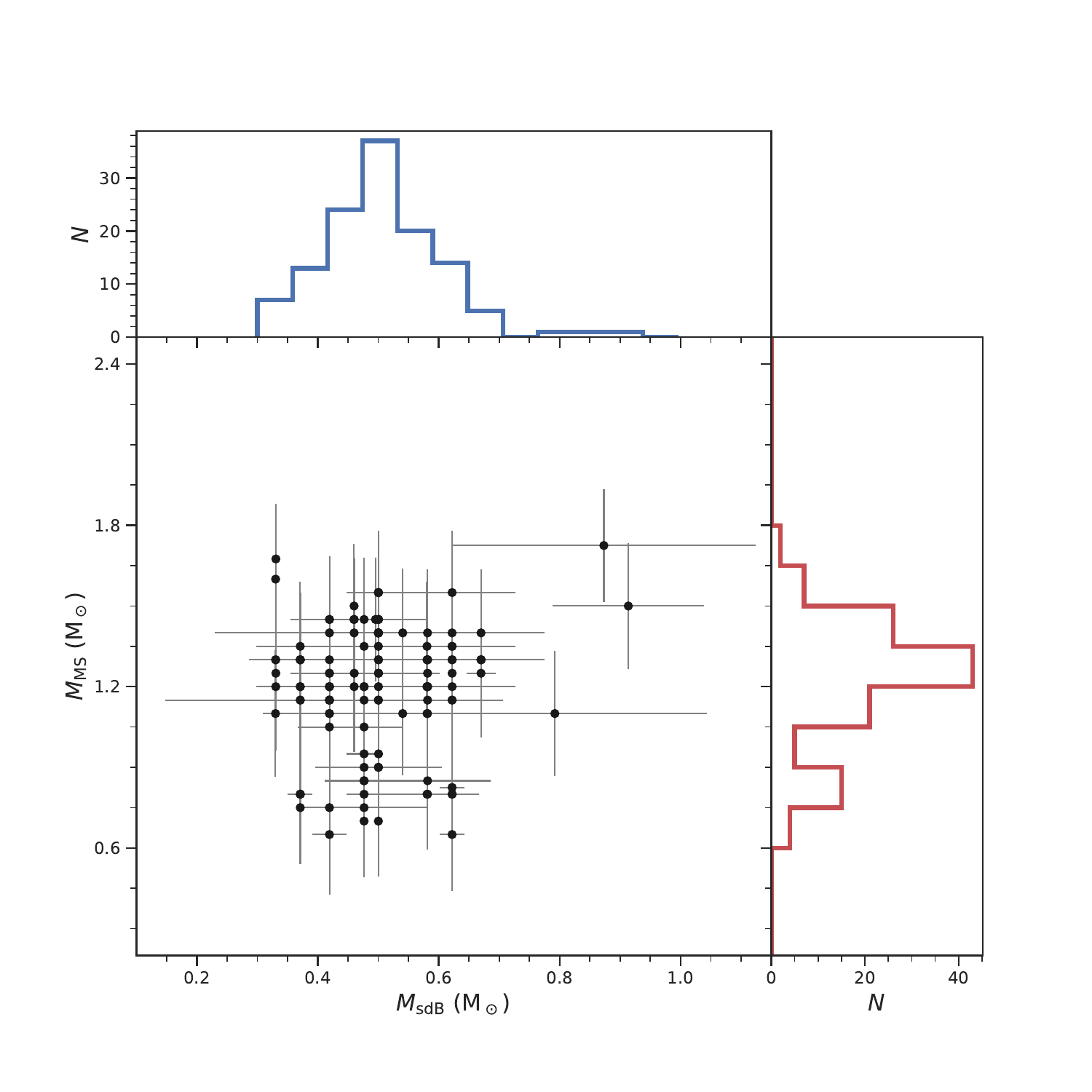}
\caption{Two-dimensional distribution of stellar masses for the sdB and MS components in our sample of 123 composite-spectrum binaries. The x-axis shows the sdB mass ($M_{\rm sdB}$) and the y-axis shows the MS companion mass ($M_{\rm MS}$), both in solar masses. }
\label{fig:mass_distribution}
\end{figure}

The distributions of stellar masses in binary systems offer key insights into the physical outcomes of binary interaction and the diversity of formation channels for hot subdwarf stars. Figure~\ref{fig:mass_distribution} presents the two-dimensional distribution of the masses of the hot subdwarfs ($M_{\rm sdB}$) and their MS companions ($M_{\rm MS}$) in our sample of 123 composite-spectrum binaries.
The sdB masses, derived from spectroscopic $T_{\rm eff}$ and $\log g$ values using the evolutionary tracks of \citet{zhangxianfei2009}, are tightly clustered between $0.40$ and $0.60\,{\rm M}_\odot$, with a clear peak near the canonical value of $0.50\,{\rm M}_\odot$ expected for core helium-burning stars. This narrow distribution supports the interpretation that the majority of sdBs in our sample are products of binary interactions that remove the hydrogen envelope near the tip of the red giant branch.

In contrast, the MS companion masses span a broader range from approximately $0.6$ to $1.9\,{\rm M}_\odot$, with most systems concentrated between $1.0$ and $1.4\,{\rm M}_\odot$. These values are consistent with late-F to G-type main-sequence stars. The underrepresentation of systems with lower-mass companions ($M_{\rm MS} < 0.8\,{\rm M}_\odot$) likely reflects observational selection effects in the study of composite systems: less massive MS stars contribute a smaller fraction of the optical flux in composite spectra, making their signatures more difficult to detect and decompose in low-resolution data.
The observed MS mass distribution shows a possible bimodal structure, consistent with a scenario in which a fraction of systems formed via the CE channel and the rest via the RLOF channel. In particular, post-CE systems with lower-mass companions are predicted to occupy a distinct region of parameter space, yet these appear underrepresented \citep{Han2003, Rodriguez-Segovia2025MNRAS_sdBpopulation}. This may be due to their intrinsically lower brightness in the optical or to the challenge of detecting the cool companion in spectroscopic surveys such as LAMOST-LRS. As such, the dominance of intermediate-mass MS companions in our sample may suggest that the stable RLOF channel plays a leading role in producing the composite-spectrum sdB binaries detected here.

In summary, the combination of a relatively narrow sdB mass distribution and a broader, but observationally biased, distribution of MS companion masses is consistent with theoretical expectations for sdB binaries formed through binary interaction. However, the apparent absence of CE-formed systems in the mass distribution highlights the influence of selection effects in spectroscopic surveys and suggests that identifying lower-mass companions may require methods beyond those based on composite spectra alone.

\subsection{Mass Ratio--Period Relation}
\label{sec:p_q}

\begin{figure}
\centering
\includegraphics[width=0.9\textwidth]{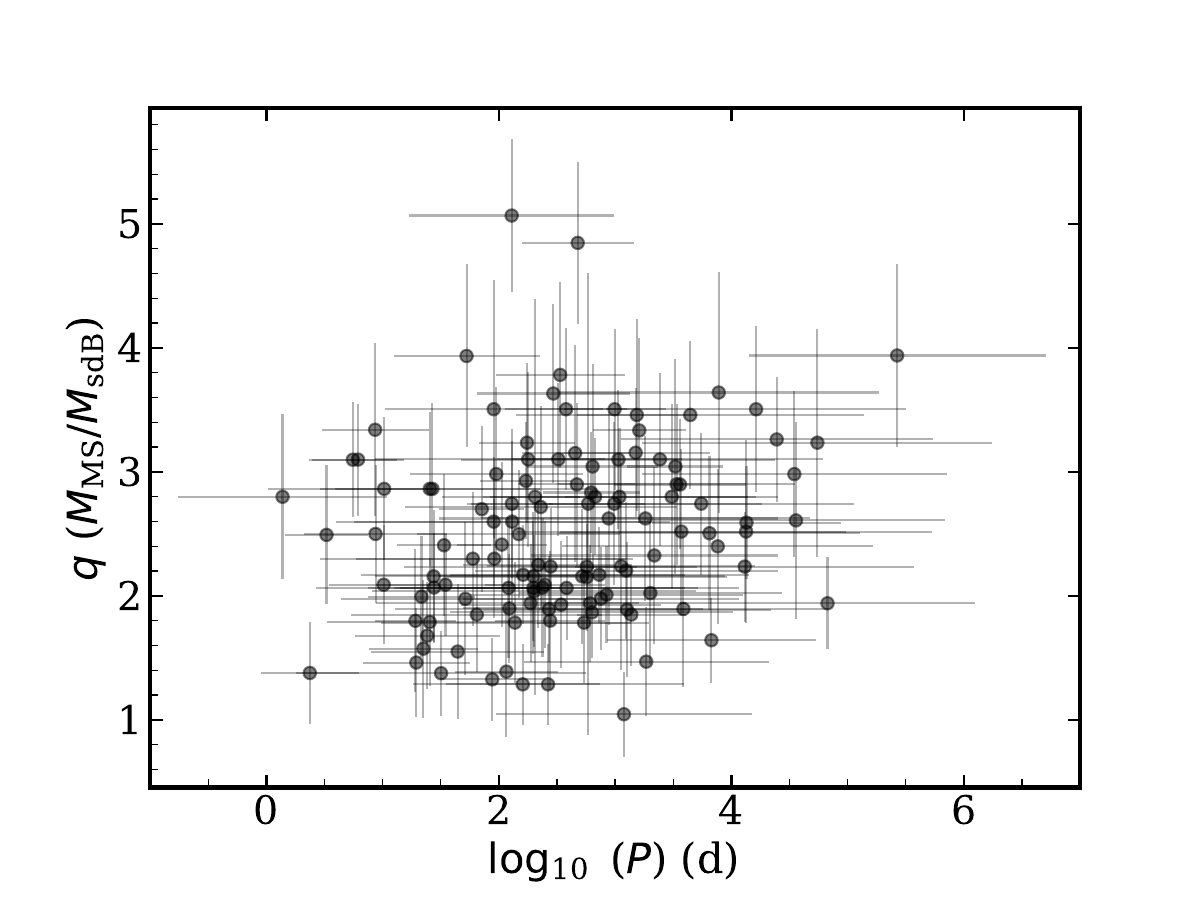}
\caption{
Mass ratio ($q = M_{\mathrm{MS}} / M_{\mathrm{sdB}}$) versus estimated orbital period ($P_{\mathrm{orb}}$) for 123 composite-spectrum sdB+MS binaries.
}
\label{fig:P_q}
\end{figure}

The relationship between orbital period and mass ratio ($q = M_{\mathrm{MS}} / M_{\mathrm{sdB}}$) serves as an important diagnostic of binary evolution, particularly in systems shaped by stable RLOF or CE interactions.
Figure~\ref{fig:P_q} presents the mass ratio $q$ plotted against the statistically inferred orbital period $P_{\mathrm{orb}}$. 
Unlike the strong anti-correlation between $P$ and $q$ reported by \citet{Vos2019MNRAS}, our data does not reveal a statistically significant trend. The points appear scattered without a clear monotonic behavior. 
As shown in Section~\ref{sec:orbital_period}, orbital periods inferred from single-epoch $\Delta{RV}$ measurements suffer from severe degeneracies with respect to orbital phase and inclination. Nevertheless, the forward-modeling tests demonstrate that the dominant period regime at $P \sim 10^{2}$–$10^{3}$~days is robustly recovered at the population level, largely independent of the assumed intrinsic period distribution. We therefore restrict our analysis to this statistically reliable regime.
Although a very weak positive trend may be visually suggested at long orbital periods ($100 < P < 1600$ days), its statistical significance is marginal. A Spearman rank correlation test yields a coefficient of $\rho = 0.20$ with a $p$-value of 0.16, indicating no statistically significant correlation between orbital period and mass ratio.

\section{Discussion and Evolutionary Implications} \label{sec:discussion}

\subsection{Wilson Relation and Systemic Velocity Consistency}

For a subset of our sample with multiple-epoch RV measurements available from LAMOST-LRS DR 12, we investigated whether dynamical methods can provide independent estimates of the systemic velocity. 
Specifically, we applied the classical $RV$--$RV$ relation introduced by \citet{Wilson1941ApJ_binary_q}, which models the radial velocities of two components in a binary system as:

\begin{equation}
RV_{\rm MS} = \gamma (1 + q) - q \cdot RV_{\rm sdB},
\label{equa:gamma1}
\end{equation}

where $RV_{\rm MS}$ and $RV_{\rm sdB}$ are the observed radial velocities of the MS and sdB components, $\gamma$ is the systemic velocity, and $q = M_{\rm MS} / M_{\rm sdB}$ is the mass ratio.
For each system with at least two epochs of RV measurements, we used linear regression to fit this equation and derive an estimate of the systemic velocity, hereafter denoted as $\gamma_{\rm Wilson}$.

\begin{figure}
\centering
\includegraphics[width=0.8\textwidth]{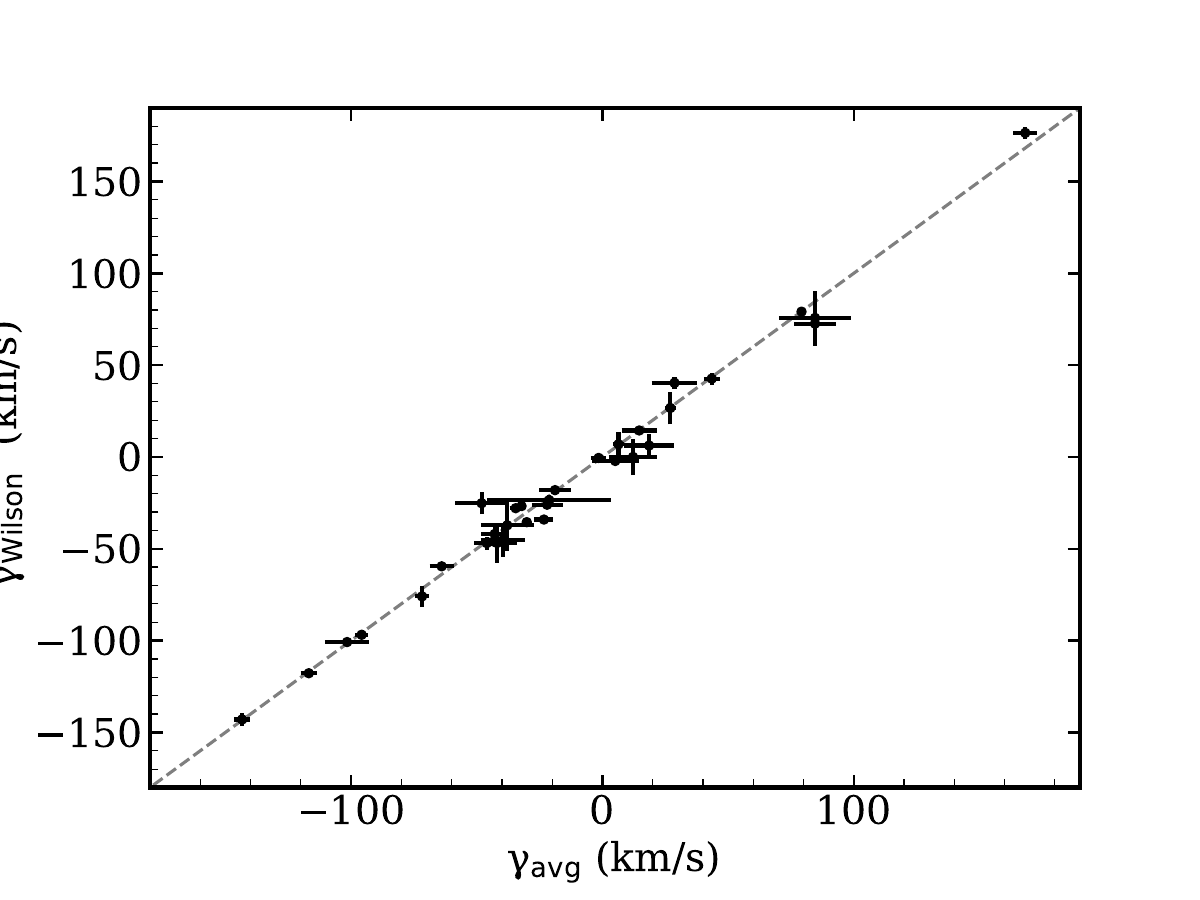}
\caption{
Comparison of systemic velocities for composite-spectrum sdB+MS binaries with multiple-epoch LAMOST observations. 
$\gamma_{\rm avg}$ is computed from the mass-weighted average of the instantaneous radial velocities of the two components, while $\gamma_{\rm Wilson}$ is derived from the Wilson RV-RV relation.
The agreement between the two estimates indicates internal consistency of the measured radial velocities and supports the binary nature of these systems.
No constraint on the mass ratio is implied.
}

\label{fig:gamma_comparison}
\end{figure}

To assess the reliability of the measured RVs, we compared $\gamma_{\rm Wilson}$ with an independent estimate based on model-derived stellar masses and instantaneous RVs. Specifically, the systemic velocity can also be estimated from:

\begin{equation}
\gamma_{\rm avg} =
\frac{M_{\rm sdB} \cdot RV_{\rm sdB} + M_{\rm MS} \cdot RV_{\rm MS}}
     {M_{\rm sdB} + M_{\rm MS}},
\label{equa:gamma2}
\end{equation}

where $M_{\rm sdB}$ and $M_{\rm MS}$ are obtained from evolutionary models as described in Section~\ref{sec:parameters}. We find that for the systems with multiple RV measurements, the systemic velocities derived using Equations~(\ref{equa:gamma1}) and~(\ref{equa:gamma2}) are in good agreement within their respective uncertainties (Figure~\ref{fig:gamma_comparison}). 
This agreement indicates that the RVs of both components are mutually consistent and follow the expectation for binary orbital motion, providing additional support that these systems are genuine binaries rather than chance line-of-sight alignments.

Although the intercept of the $RV$-$RV$ relation (i.e., $\gamma$) is relatively robust, the slope, which determines the mass ratio, is highly sensitive to measurement uncertainties. In particular, the radial velocities of the MS companions are more susceptible to error due to weaker spectral features and lower signal-to-noise ratios in the red spectral region. Furthermore, the limited number of spectroscopic epochs, typically only two or three per system, provides insufficient phase coverage to constrain the full orbital motion. As a result, the mass ratios inferred from the Wilson relation exhibit large scatter and are not used to validate or refine the mass ratios estimated from evolutionary models. In summary, the Wilson analysis is employed here solely as a consistency check on the systemic velocity and on the internal coherence of the measured radial velocities. Robust dynamical constraints on mass ratios will require future multi-epoch spectroscopic observations with adequate phase coverage.

\subsection{Implications for Binary Evolution Models}

The orbital period and mass ratio of sdB binaries provide valuable diagnostics for understanding their evolutionary histories. In particular, systems formed via the CE ejection channel are typically predicted to have short orbital periods ($P \lesssim 10$ days), while those formed through stable RLOF can retain wider separations, often with periods exceeding hundreds or even thousands of days \citep{Han2003, Kupfer2015}. 
The statistically inferred orbital periods in our sample span this full range (Section~\ref{sec:orbital_period}). Despite the large uncertainties affecting individual systems, the population-level distribution remains clearly concentrated toward long orbital periods, spanning roughly a few tens to several thousand days. Such long-period configurations are broadly consistent with expectations for binaries formed through stable Roche-lobe overflow, although we emphasize that this conclusion applies only at the statistical level and not to individual binaries.

Our sample is inherently biased toward composite-spectrum sdB+MS systems with detectable main-sequence companions of spectral type F, G, or K.
Consequently, the survey is less sensitive to systems with very high‑mass companions such as O or B stars, or with very low‑mass companions such as M dwarfs, and therefore preferentially selects binaries with intermediate‑mass companions and wider orbits.
This observational bias naturally enhances the contribution of systems formed through stable RLOF in the observed population.
In such systems, the comparatively high companion masses are expected to promote dynamically stable mass transfer during the red-giant phase of the sdB progenitor, reducing the likelihood of a common-envelope event. Consequently, the observed orbital period and mass ratio distributions reflect not only the underlying binary evolution physics but also strong selection effects inherent to composite-spectrum samples.

Future multi-epoch spectroscopic observations will be essential to determine complete orbital solutions, enabling a more robust comparison with theoretical models and refining our understanding of the formation channels of hot subdwarf binaries. 
In addition, detailed chemical abundance studies of the MS companions, such as the work of \citet{Molina2022A&A_sdB}, will provide complementary constraints on the evolutionary history by probing signatures of past mass transfer.

\section{Summary and Conclusions} \label{sec:summary}

In this work, we present a comprehensive analysis of 123 composite-spectrum hot subdwarf binaries with main-sequence companions identified from the LAMOST-LRS DR 8. This sample, originally compiled by \citet{LeiZhenxin2023ApJ}, represents one of the largest homogeneous collections of such systems identified to date through spectroscopic decomposition. By combining atmospheric parameters from spectral fitting, stellar masses and radii estimated from evolutionary models, and radial velocity measurements from LAMOST spectra, we explore the physical and orbital properties of these binaries and their implications for binary evolution.

Atmospheric parameters for both the sdB and MS components were adopted from the spectral decomposition by \citet{LeiZhenxin2023ApJ}. We refined the stellar masses and radii using theoretical evolutionary tracks, adopting the hot subdwarf models from \citet{zhangxianfei2009} and the ZAMS models from PARSEC for the main sequence stars.
Radial velocities for both components were measured independently using cross-correlation with synthetic templates, allowing us to estimate instantaneous velocity separations.
Given the limited number of spectroscopic epochs for most systems, direct orbital solutions are not feasible. Instead, we estimated orbital periods in a statistical sense using Monte Carlo simulations. This method assumes circular orbits and uses single-epoch RV separations, combined with stellar masses, to infer a distribution of possible periods for each system. 
The inferred orbital-period distributions span a wide range, from a few days to several thousand days. At the population level, the distribution is concentrated toward long orbital periods, extending from a few tens to several thousand days, consistent with wide systems likely formed through stable Roche-lobe overflow. Although individual orbital periods remain poorly constrained, this dominant regime is robust against the strong degeneracies inherent to single-epoch RV measurements. Future multi-epoch spectroscopic monitoring will be required to more tightly constrain the intrinsic period distribution. Rather than over-interpreting sparse RV information, our approach provides a statistically motivated framework for extracting population-level constraints from single-epoch observations, thereby maximizing the scientific return of such datasets.

We examined the mass distributions of the sdB and MS components and found that most sdB stars have masses tightly clustered around $0.50\,{\rm M}_\odot$, consistent with core helium-burning objects, while MS companions span a wider mass range of $0.6$–$1.9\,{\rm M}_\odot$.
We also investigated the mass ratio–period relation and found no statistically significant correlation.
Finally, we discussed the implications of our results for models of binary evolution.
Since our analysis focuses on composite spectrum sdB binaries, mainly sdB+FGK systems, our approach naturally favors the detection of long period binaries that are products of stable Roche lobe overflow, providing one of the largest homogeneous samples of such systems to date. Future multi-epoch spectroscopic monitoring will be essential to obtain complete orbital solutions and to more clearly distinguish the evolutionary channels of composite spectrum sdB binaries. In addition, detailed abundance analyses of the companion stars will help trace the mass transfer history and place tighter constraints on the formation pathways of hot subdwarf binaries within the broader context of binary stellar evolution.

\begin{acknowledgments}

We thank the anonymous referee for his/her valuable comments. This work is supported by National Natural Science Foundation of China (Grant Nos. 12125303, 12288102, 12090040/3, 12403040, 12422305, 12333008, 12525304), National Key R$\&$D Program of China (Grant No. 2021YFA1600401/3), the International Centre of Supernovae (No. 202201BC070003), Yunnan Key Laboratory (No. 202302AN360001) and the Yunnan Revitalization Talent Support Program-Science \& Technology Champion Project (No. 202305AB350003) and Young Talent Project. We also acknowledge the science research grant from the China Manned Space Project with No.CMS-CSST-2021-A10 and the CAS Light of West of China Program. This work has been supported by the New Cornerstone Science Foundation through the XPLORER PRIZE and by the Strategic Priority Research Program of the Chinese Academy of Sciences (grant Nos. XDB1160000 / XDB1160200). Jiao Li is supported by the Yunnan Fundamental Research Projects (YFRP) grant No. 202501CF070016 and the Young Talent Project of Yunnan Revitalization Talent Support Program.

Guoshoujing Telescope (LAMOST) is a National Major Scientific Project built by the Chinese Academy of Sciences. Funding for the project has been provided by the National Development and Reform Commission. LAMOST is operated and managed by the National Astronomical Observatories, Chinese Academy of Sciences. 

This work has made use of data from the European Space Agency (ESA) mission Gaia (https://www.cosmos.esa.int/gaia), processed by the Gaia Data Processing and Analysis Consortium (DPAC, https://www.cosmos.esa.int/web/gaia/dpac/consortium). Funding for the DPAC has been provided by national institutions, in particular the institutions participating in the Gaia Multilateral Agreement.

We acknowledge the support of the staff of the Lijiang 2.4m telescope. 
Funding for the telescope has been provided by Chinese Academy of Sciences and the People's Government of Yunnan Province.
\end{acknowledgments}

\section*{Data Availability}

The LAMOST DR8 spectra used in this study are publicly available through the LAMOST archive. 

%\begin{contribution}
%%This section gives authors the space to recognize author contributions. The text inside this environment is NOT counted towards the total word quanta. At a minimum, manuscripts are expected to include this text:

%%All authors contributed equally to the Terra Mater collaboration.

%% But authors are expected to provide more specific details, e.g. 
%%
%%SC was responsible for writing and submitting the manuscript.
%%WWM came up with the initial research concept and edited the manuscript.
%%OTS obtained the funding and edited the manuscript.
%%EBF provided the formal analysis and validation. He also edited the manuscript.
%%GEH Supervised the undergraduates, wrote the software and administers the project github and Zenodo repositories.
%%
%% Authors can use the Contributor Role Taxonomy (CRediT) at
%% for ideas on how write a good statement tailored to their needs.

%\end{contribution}

\facilities{LAMOST, Gaia}

\software{astropy \citep{astropy:2022},  
          PARSEC \citep{Bressan2012MNRAS_PARSEC, Chenyang2014MNRAS_PARSEC,Chenyang2015_PARSEC},
          laspec \citep{Zhangbo2020ApJS, Zhangbo2021ApJS}}


%% For this sample we use BibTeX plus aasjournalv7.bst to generate the

\bibliography{ref}{}
\bibliographystyle{aasjournalv7}



\end{document}